\documentclass[lettersize,journal]{IEEEtran}
\usepackage{amsmath,amsfonts}
\usepackage{algorithmic}
\usepackage{algorithm}
\usepackage{array}
\usepackage[caption=false,font=normalsize,labelfont=sf,textfont=sf]{subfig}
\usepackage{textcomp}
\usepackage{stfloats}
\usepackage{url}
\usepackage{verbatim}
\usepackage{graphicx}
\usepackage{cite}
\usepackage[numbers,sort&compress]{natbib}
\usepackage{hyperref}
\usepackage{multirow}
\usepackage{tabularx}
\usepackage{array} 
 \usepackage{booktabs}
 \usepackage{makecell}
 \usepackage[table,xcdraw]{xcolor}
\usepackage{longtable}
\usepackage{booktabs}
\usepackage{amssymb}
\usepackage{bbding}
\usepackage{pifont}
\usepackage{wasysym}
\usepackage{utfsym}
\usepackage[figuresright]{rotating}
\usepackage{fontawesome}
 \usepackage[section]{placeins}
\everymath{\displaystyle}
\begin{document}

\title{Manipulated Regions Localization For Partially Deepfake Audio: A Survey}

\author{Jiayi He, Jiangyan Yi~\IEEEmembership{Member, IEEE,}, Jianhua Tao*~\IEEEmembership{Senior Member, IEEE}, Siding Zeng, and Hao Gu
\thanks{Jiayi He is with the State Key Laboratory of Multi-modal Artificial Intelligence Systems, Institute of Automation, Chinese Academy of Sciences. E-mail: jiayi.he@ia.ac.cn}
\thanks{Jiangyan Yi is with the Department of Automation, Tsinghua University, Beijing, China. E-mail: yijy@tsinghua.edu.cn}
\thanks{Jianhua Tao is with the Department of Automation and Beijing National Research Center for Information Science and Technology, Tsinghua University, Beijing, China. E-mail: jhtao@tsinghua.edu.cn (*Corresponding author)}
\thanks{Siding Zeng is with University of Chinese Academy of Sciences and the State Key Laboratory of Multi-modal Artificial Intelligence Systems, Institute of Automation, Chinese Academy of Sciences. E-mail: zengsiding2023@ia.ac.cn}
\thanks{Hao Gu is with the State Key Laboratory of Multi-modal Artificial Intelligence Systems, Institute of Automation, Chinese Academy of Sciences. E-mail: guhao2022@ia.ac.cn}}



\maketitle

\begin{abstract}
With the development of audio deepfake techniques, attacks with partially deepfake audio are beginning to rise. Compared to fully deepfake, it is much harder to be identified by the detector due to the partially cryptic manipulation, resulting in higher security risks. Although some studies have been launched, there is no comprehensive review to systematically introduce the current situations and development trends for addressing this issue. Thus, in this survey, we are the first to outline a systematic introduction for partially deepfake audio manipulated region localization tasks, including the fundamentals, branches of existing methods, current limitations and potential trends, providing a revealing insight into this scope.
\end{abstract}


\section{Introduction}
\IEEEPARstart{W}{ith} the rapid development of artificial intelligence generated content (AIGC) techniques, significant improvements have been made in the naturalness, realism, and diversity of the synthetic audio. However, at the meanwhile, the misuse of the advanced technology may also poses a serious threat to social security, cyber security, and privacy security. In order to defend against these issues, deepfake audio detection had raised the attention in the past few years. To date, many effective countermeasures (CMs) have emerged\cite{9747766,10096837,10094704,WU2015130,kamble2020advances,tan2021survey,mittal2022automatic,almutairi2022review,yi2023audio,xie2024codecfakedatasetcountermeasuresuniversally}, and the performance of some models evaluated through equal error rate (EER) are reported to be less than 1\%\cite{9747766,xie2024codecfakedatasetcountermeasuresuniversally}, indicating significant success in defending against fully deepfake audio attacks. 
\begin{figure}
    \centering
    \includegraphics[width=1\linewidth]{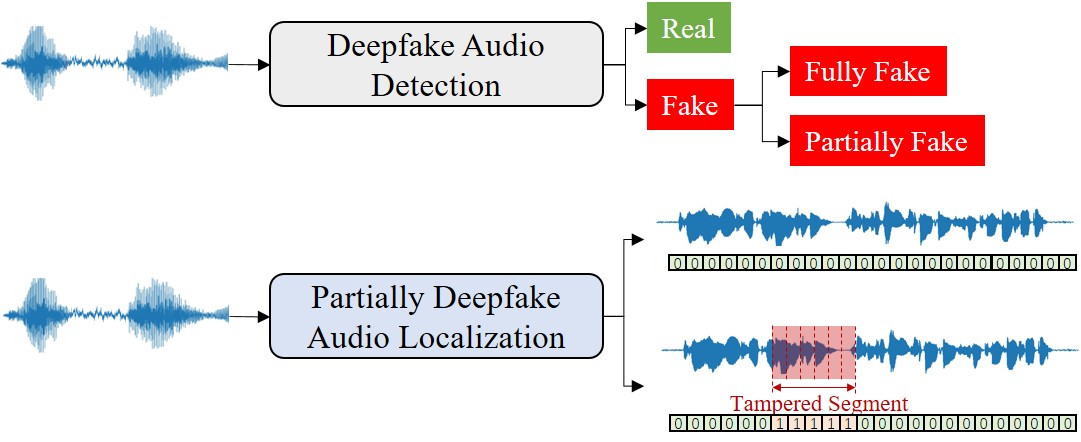}
    \caption{The differences between detection tasks and localization tasks.}
    \label{fig:diff}
\end{figure}
However, with each step forward, new challenges are emerged. Partially deepfake audio attacks, a more covert way of spoofing, have attracted another round of attention. The partially deepfake audio usually combines real and fake audio clips or another real clips from other corpus, increasing the complexity and difficulty of recognizing the attacks. Existing research shows that both humans and machines can be easily deceived by partially deepfake audio\cite{jcp5010006}. To cope with this new challenge, in recent years, some fundamental facilities, such as diverse datasets, and competitions are launched to attract the attention in the community\cite{2021An,yi21_interspeech,9956134,conf/icassp/YiFTNMWWTBFLWZY22,yi2023add,10888070} and some CMs are proposed to locate the partially manipulations\cite{10037855,9956134,xie2024efficient,liu2023transsionadd,cai2023waveform,9746162,Wang2022SyntheticVD,zhang2021multi,martin2023vicomtech,cai2023dku,inbook,li2023convolutional,rahman2022detecting,kumar2021speech,li2023multi,zhu2023local,10889913,10890470,zhong2024enhancing}. Up to now, for the commonly used datasets, the best performance reported was a frame-level EER of 3.58\%\cite{zhong2024enhancing} on PartialSpoof datasets\cite{2021An} and a segment-level F1-score of 0.7397\cite{10889913} on ADD2023Track2 datasets\cite{yi2023add}. In PartialSpoof datasets, all manipulated regions are generated by TTS/VC. However, with the development of spoofing techniques, genuine clips are also used for tampering. The ADD2023Track2 datasets takes the situation of 'truth for truth' into consideration. Besides, additional noise, format conversion and the smoothing processing on spliced traces were done. Moreover, with the widely application of large language model (LLM), advanced techniques for local feature matching and seamless stitching are bound to arrive, which will pose new challenges for partially deepfake audio localization tasks. Therefore, in order to better defend against it in the near future, we urgently need a comprehensive review to help understand the current situations, including existing outperforming CMs and the development trends of this issue.

Thus, in this survey, we aim to provide a comprehensive overview of the current state in this scope, summarizing and comparing existing methodologies, and highlighting their respective strengths and weaknesses. Also, it is organized to guide future research directions and foster technological advancements by identifying gaps and challenges in the current research that remain. Ultimately, this survey aims to enhance researchers' understanding and raise community awareness of manipulated regions localization tasks for partially deepfake audio. Additionally, we also hope that it can become a guide for beginners in this research scope. The contributions of this survey are presented as following: 
\begin{itemize}
    \item This is the first comprehensive survey focusing on partially deepfake audio manipulated regions localization tasks. 
    \item This survey provide a comprehensive summary, including the difficulty and specificity of the task, as well as the categories and the performance comparison of existing methods, which is particularly helpful for enhancing the understanding and raising community awareness.
    \item Specifically, the best performance for some commonly used diverse datasets are collected. Based on the current performance, the remaining challenges and limitations are discussed. At the meanwhile, some potential development trends are also discussed.
\end{itemize}

\section{Definition and Characteristics of Partially Deepfake Audio Manipulation Localization}
\subsection{Definition of Partially Deepfake Audio}
\begin{figure*}
    \centering
    \includegraphics[width=1\linewidth]{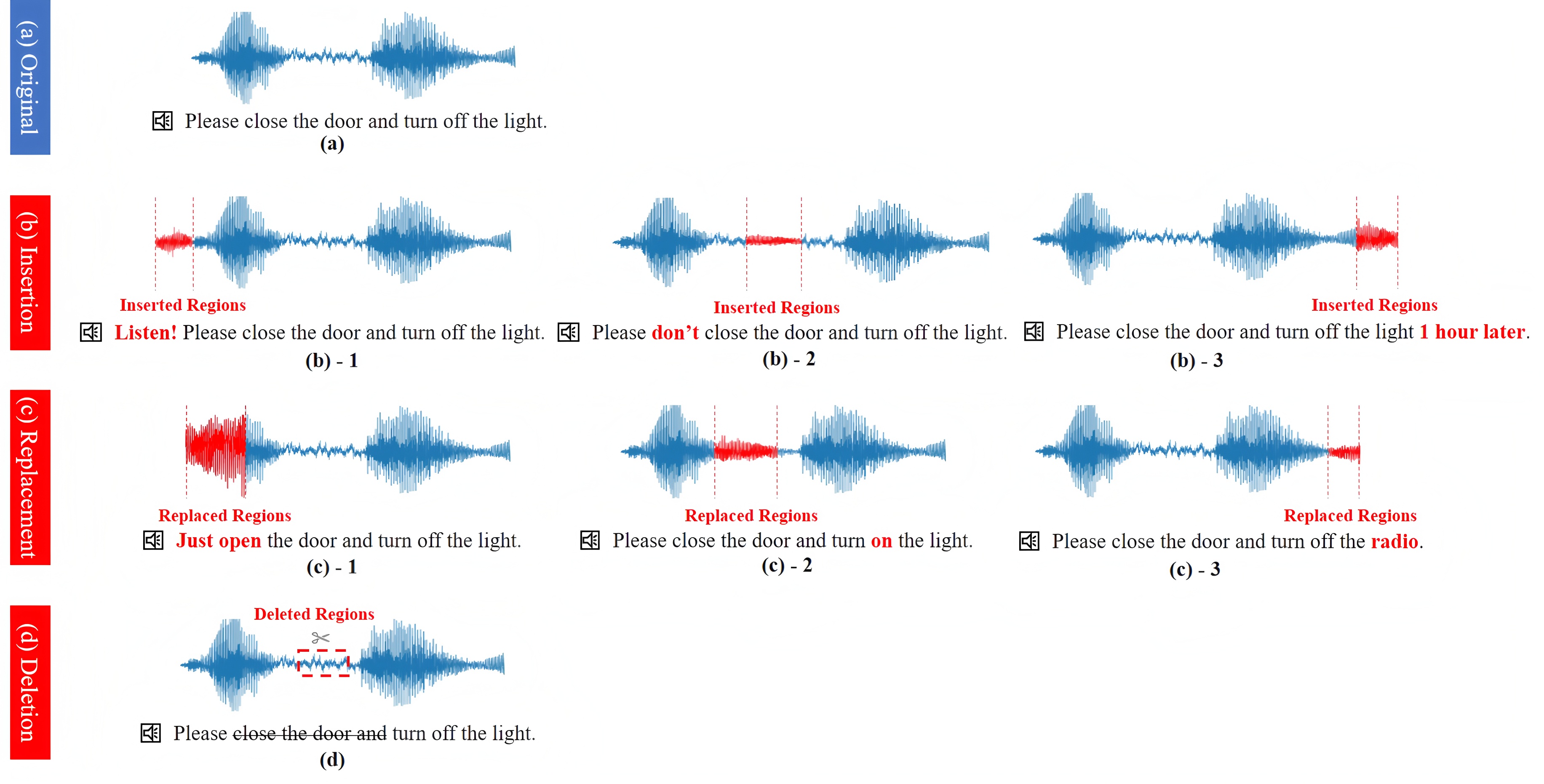}
    \caption{Original bona fide audio and different types of partially deepfake audio: (a) Original bona fide audio; (b) Insertion at the beginning, middle, and the end of the utterance respectively; (c) Replacement at the beginning, middle, and the end of the utterance respectively; (d) Deletion at the middle of the utterance.}
    \label{fig:types}
\end{figure*}

Partially deepfake audio refers to the audio with partial manipulations. The source of manipulated clips may be either synthetic or bona fide. The splicing method can include insertion, replacement, deletion, etc. Fig.\ref{fig:types} shows the illustration of the different types. From the illustration, it can be seen that the meaning or tone of utterances can be changed by simply manipulating certain regions of it.
\subsection{Differences from Fully Deepfake Audio}
\subsubsection{Process of Generation}
The fully deepfake audio is usually entirely synthesized via text-to-speech (TTS)\cite{10.1016/j.specom.2014.10.005}, voice conversion (VC)\cite{10.1016/j.specom.2014.10.005,10.1016/j.specom.2021.11.006}, Emotion Fake\cite{zhao2023emofake}, Scene Fake\cite{Yi2022SceneFakeAI}, etc., focusing on the fidelity and naturalness of the entire audio. While, for partially deepfake audio, only a few clips in a genuine audio are manipulated. The entire process generally includes three steps: selecting the splicing position, preparing the manipulated segments, and splicing the manipulated segments\cite{yi21_interspeech}. Its main concern is to ensure that the manipulated clips are highly consistent with real clips and avoiding leaving stitching marks.
In summary, partially deepfake audio generation focuses on the high-quality operations of local substitution and seamless stitching, while fully deepfake audio generation emphasizes on global fidelity and naturalness.
\subsubsection{Purpose of Spoofing}
Fully deepfake audio spoofing is usually used to generate the voice of the target objects to achieve sound deception. While partially deepfake audio spoofing tends to change the expression of the original voice command by editing a few key words to implement the specific intent tampering.
\subsection{Differences from Detection Tasks}
Deepfake audio detection tasks mainly focus on the binary classification of genuine audio and fake or partially fake audio, and provides sentence-level labels, focusing more on the absolute authenticity of global features. While partially deepfake audio manipulation localization tasks emphasize to identify the manipulated regions in the audio and provide segment-level identification by discovering the local inconsistency of the audio itself (See Fig.\ref{fig:diff}). In special circumstances, localization tasks can be considered as segment-level detection tasks when and only when the manipulated segment is fake.

\section{Related Work}
\subsection{Fundamental Facilities}

\subsubsection{Datasets}
To date, there have been several established datasets for partially deepfake audio localization tasks. The information is shown in Table \ref{tab:dataset}.
\begin{table*}[]
\centering
\caption{The details of datasets for partially deepfake audio manipulated regions localization tasks.}
\label{tab:dataset}
\begin{tabular}{ccccccccccc}

\toprule[2pt]
\rowcolor[HTML]{EFEFEF} 
\cellcolor[HTML]{EFEFEF}                       & \cellcolor[HTML]{EFEFEF}                       & \cellcolor[HTML]{EFEFEF}                          & \cellcolor[HTML]{EFEFEF}                           & \cellcolor[HTML]{EFEFEF}                           & \cellcolor[HTML]{EFEFEF}                                                                                           & \cellcolor[HTML]{EFEFEF}                                                                                  & \cellcolor[HTML]{EFEFEF}                         & \multicolumn{3}{c}{\cellcolor[HTML]{EFEFEF}\#Utterances}                                                                     \\ \cline{9-11} 
\rowcolor[HTML]{EFEFEF} 
\multirow{-2}{*}{\cellcolor[HTML]{EFEFEF}Ref.} & \multirow{-2}{*}{\cellcolor[HTML]{EFEFEF}Year} & \multirow{-2}{*}{\cellcolor[HTML]{EFEFEF}Dataset} & \multirow{-2}{*}{\cellcolor[HTML]{EFEFEF}Language} & \multirow{-2}{*}{\cellcolor[HTML]{EFEFEF}Modality} & \multirow{-2}{*}{\cellcolor[HTML]{EFEFEF}\begin{tabular}[c]{@{}c@{}}The Type of \\ Manipulated Clips\end{tabular}} & \multirow{-2}{*}{\cellcolor[HTML]{EFEFEF}\begin{tabular}[c]{@{}c@{}}Manipulation \\ Methods\end{tabular}} & \multirow{-2}{*}{\cellcolor[HTML]{EFEFEF}Access} & Real    & \begin{tabular}[c]{@{}c@{}}Fully \\ Fake\end{tabular} & \begin{tabular}[c]{@{}c@{}}Partially \\ Fake\end{tabular} \\ \hline
\cite{2021An,zhang_2022_5766198}               & 2021                                           & PartialSpoof                              & English                                            & Audio                                              & Generated                                                                                                          & TTS/VC                                                                                                    & Public                                           & 12,483  & 0                                                     & 108,978                                                   \\
\cite{yi21_interspeech}                        & 2021                                           & HAD                                               & Chinese                                            & Audio                                              & Generated                                                                                                          & TTS                                                                                                       & Public                                           & 53,612  & 53,612                                                & 53,612                                                    \\
\cite{conf/icassp/YiFTNMWWTBFLWZY22}           & 2022                                           & ADD2022Track2                             & Chinese                                            & Audio                                              & Generated/Real                                                                                                     & TTS/Real                                                                                                       & Public                                           & 5,319   & 45,367                                                & 1,052                                                     \\
\cite{9956134}                                 & 2022                                           & Psynd                                             & English                                            & Audio                                              & Generated                                                                                                          & TTS                                                                                                       & Restrict                                         & -       & -                                                     & -                                                         \\
\cite{10034605}                                & 2022                                           & LAV-DF                                            & Multilingual                                            & Audio/Video                                        & Generated                                                                                                          & TTS                                                                                                       & Public                                           & 36,431  & 0                                                     & 99,873                                                    \\
\cite{yi2023add}                               & 2023                                           & ADD2023Track2                             & Chinese                                            & Audio                                              & Generated/Real                                                                                                     & TTS/Real                                                                                                       & Public                                           & 55,467  & 1,618                                                 & 63,831                                                    \\
\cite{10.1145/3664647.3680795}                 & 2024                                           & AV-Deepfake1M                                     & Multilingual                                            & Audio/Video                                        & Generated                                                                                                          & TTS                                                                                                       & Public                                           & 286,721 & 0                                                     & 860,039                                                   \\
\cite{10888070}                                & 2025                                           & LlamaPartialSpoof                                 & English                                            & Audio                                              & Generated                                                                                                          & TTS                                                                                                       & Public                                           & 10,573  & 33,461                                                & 32,194                                                    \\ 
\cite{example}                 & 2025                                           & AV-Deepfake1M++                                     & Multilingual                                            & Audio/Video                                        & Generated                                                                                                          & TTS                                                                                                       & Restrict                                           & - & -                                                    & -                                                   \\\bottomrule[2pt]
\end{tabular}
\end{table*}
\begin{itemize}
    \item \textbf{PartialSpoof Dataset}\footnote{PartialSpoof: \url{https://zenodo.org/records/5766198}}\cite{2021An,zhang_2022_5766198}. This is the first English dataset proposed to focus on partially deepfake audio. It is built based on ASVspoof 2019 LA database\cite{WANG2020101114,Todisco2019} and provides segment labels for various temporal resolutions\cite{Zhang_2023}. In this datasets, every partially deepfake audio is a mixture of genuine and fake clips. Segments randomly chosen from a genuine audio are replaced with spoofed one and vice versa. Both segment-level labels and sentence-level labels are provides. Segments and utterances containing one or more generated frames are labeled as \textit{spoof}, otherwise \textit{bona fide}.
    
    \item \textbf{Half-truth Dataset (HAD)}\footnote{HAD: \url{https://zenodo.org/records/10377492}}\cite{yi21_interspeech}. This is the first Chinese partially deepfake audio dataset, built based on AISHELL-3 corpus\cite{shi21c_interspeech}, consisting of partially fake, fully fake, and real audio. Compared to the PartialSpoof database, instead of randomly choosing segments to pollute the raw audio, semantic coherence and word boundaries are considered during the manipulation generation.
    
    \item \textbf{ADD2022Track2 Dataset}\footnote{ADD2022Track2 Train\&Dev: \url{https://zenodo.org/records/12188127}\\ ADD2022Track2 Adaption: \url{https://zenodo.org/records/12188083}\\      ADD2022Track2 Eval: \url{https://zenodo.org/records/12187997}}\cite{conf/icassp/YiFTNMWWTBFLWZY22}. It is designed to support the first Audio Deep synthesis Detection challenge (ADD 2022), consisting of partially fake, fully fake, and real audio. In this dataset, the partially fake audio is collected as an adaptation set, generating by manipulated the original genuine audio with real or synthesized clips. Test set consists of unseen genuine and partially fake audio, where some utterances are selected from Mandarin corpus AISHELL-1\cite{8384449}, AISHELL-3\cite{shi21c_interspeech}, and AISHELL-4\cite{Fu2021}.
    
    \item \textbf{ADD2023Track2 Dataset}\footnote{ADD2023Track2 Train\&Dev: \url{https://zenodo.org/records/12176530}\\   ADD2023Track2 Eval: \url{https://zenodo.org/records/12176904}}\cite{yi2023add}. It is designed to support the second Audio Deep synthesis Detection challenge (ADD 2023), consisting of partially fake, fully fake, and real audio. Similar to ADD2022Track2 dataset, the partially fake audio is also generating by manipulated the original genuine audio with either real or synthesized clips. The training and dev sets are also collected based on AISHELL-3. The test set includes unseen partially fake and real utterances. Different from ADD2022Track2 dataset, the training and dev sets consist of all of the three types. Besides, in test set, additional noise and format conversions were done, which significantly increased the difficulty in localization.
    
    \item \textbf{Partial Synthetic Detection dataset (Psynd)}\footnote{Psynd: \url{https://scholarbank.nus.edu.sg/handle/10635/227398}}\cite{9956134}. This dataset consists of approximately 13 hours multi-speaker English corpus, based on LibriTTS\cite{2019arXiv190402882Z}, and the fake segments are injected into real utterances.
    
    \item \textbf{Localized Audio Visual DeepFake Dataset(LAV-DF)}\footnote{LAV-DF: \url{https://huggingface.co/datasets/ControlNet/LAV-DF}}\cite{10034605}. This is the first large audio-visual deepfake dataset in manipulation localization tasks. The manipulation is rule-based and content-driven. The manipulation strategy is to replace strategic words with their antonyms, which leads to a significant change in the sentiment of the statement. In this dataset, the audio is extracted from video, and the real videos are sourced from the VoxCeleb2 dataset\cite{chung2018voxceleb2}. The partial fake is triggered by transcript manipulation, and the corresponding partially fake audio is generated by SV2TTS\cite{jia2018transfer}.
    
    \item \textbf{AV-Deepfake1M Dataset}\footnote{AV-Deepfake1M: \url{https://huggingface.co/datasets/ControlNet/AV-Deepfake1M}}\cite{10.1145/3664647.3680795}. It is a further step of content-driven audio-visual deepfake dataset for manipulation localization tasks. Different from LAV-DF, it employed ChatGPT for altering the real transcripts, ensuring the diversity and context consistent. It includes two additional challenging manipulation strategies, deletion and insertion, more than replacement. Besides, VITS\cite{kim2021conditional} and YourTTS\cite{casanova2022yourtts} are employed to generated the fully fake and partially fake audio. Its size is nearly ten times that of LAV-DF. Recently, AV-Deepfake1M++ dataset is released\cite{example}, containing over 2 million samples

    \item \textbf{LlamaPartialSpoof Dataset}\footnote{LlamaPartialSpoof: \url{https://huggingface.co/datasets/HaoY0001/LlamaPartialSpoof}}\cite{10888070}. It is a content-driven deepfake dataset with audio only. This dataset is designed to enhance the quality and diversity of fully fake and partially fake utterances, built based on LibriTTS. Inspired by AV-Deepfake1M, Llama-3-8B-Instruct is employed to automatically alter sentences. The difference is that, in this dataset, the model is asked to change the transcript via several prompts instead of generating a series of replace, delete, and insert operations , improving the quality of manipulated transcription. Five TTS models are adopted to generate the fully fake and partially fake audio. The partially fake audio in this dataset is concatenated by real and fake segments. Post-process is done for both bona fide and the fake utterances.
\end{itemize}
\subsubsection{Evaluation metrics}
\begin{itemize}
    \item \textbf{Segment-level EER}. Zhang \textit{et al}\cite{2021An} proposed to adapt utterance-level EER to segment-level EER at first, named as point-based EER. The definition is showing below: 
    \begin{equation}
        FPR = \frac{FP}{FP+TN}
    \end{equation}
    \begin{equation}
        FNR = \frac{FN}{FN+TP}
    \end{equation}
    where FP refers to the real segments that are incorrectly detected as fake, and TN refers to the real segments that are correctly detected as real, and FN refers to the fake segments that are wrongly detected as real, and TP refers to the fake segments that are correctly detected as fake. When FPR = FNR, the common value is EER, which is widely used in the binary classification tasks.
    Then, to correct the precisions that caused by some potential misclassified regions and relieve the impact of diverse resolution, they modified it to range-based EER\cite{zhang23v_interspeech}.
    \begin{equation}
    \label{eq:eer}
    EER=\frac{FPR(\tau)+FNR(\tau)}{2},
    \end{equation}
    where  
    \begin{equation}
    \label{eq:FPR-range}
    FPR(\tau)=\frac{\sum_{i\in Hypo}\sum_{j\in Ref}\mathcal{I}(Pred<\tau)\mathcal{T}(r_i,r_j)}{Duration\quad of\quad Negtive\quad Label},
    \end{equation}
    \begin{equation}
    \label{eq:FNR-range}
    FNR(\tau)=\frac{\sum_{i\in Hypo}\sum_{j\in Ref}\mathcal{I}(Pred\geq\tau)\mathcal{T}(r_i,r_j)}{Duration\quad of\quad Positive\quad Label}.
    \end{equation}
    $\mathcal{I}(\cdot)$ denotes the indicator function that outputs 1 when the condition is true and 0 otherwise. $\mathcal{T}(\cdot)$ records the overlap of two ranges. $\tau$ is obtain by binary search algorithm as described in \cite{zhang23v_interspeech}. Instead of labeling every segments, in this metric, reference labels are given according to the boundaries and duration for manipulation regions.
    
    \item \textbf{Segment-level Precision (P), Recall (R) and F1-score ($F_1$)}. Yi \textit{et al}\cite{yi21_interspeech} proposed to use these metrics to evaluate the performance of localization accuracy, which are based on the duration of each segment.
    \begin{equation}
        P = \frac{TP}{TP+FN}
    \end{equation}
    \begin{equation}
        R = \frac{FN}{TP+FP}
    \end{equation}
    \begin{equation}
        F_1 = \frac{2PR}{P+R}
    \end{equation}
    where TP refers to the fake segments that are correctly detected as fake, and FN refers to the fake segments that are incorrectly detected as real, and FP refers to the real segments that are wrongly detected as fake.
    
    \item \textbf{The weighted sum of sentence-level Accuracy(Acc) and segment-level $F_1$}. In ADD 2023 Track 2\cite{yi2023add}, the evaluation is designed to focus on both sentence-level and segment-level performance at the same time. Thus, it is defined as a weighted sum of \textit{Sentence Accuracy} and \textit{Segment $F_1$}, as shown in Eq.\ref{eq1}.
    \begin{equation}
    \label{eq1}
    Score=0.3\times Acc+0.7\times F_1,
    \end{equation}
    where 
    \begin{equation}
    \label{eq2}
    Acc=\frac{TP+TN}{TP+TN+FP+FN},
    \end{equation}
    \begin{equation}
        F_1 = \frac{2PR}{P+R}
    \end{equation}
     where the definition of TP, FP, TN, and FN are consistent with above. It is worth noting that these statistics in the Acc related formulas are at the sentence level, while those related to $F_1$ are at the segment level.

     \item \textbf{1D-Intersection over Union (IoU)}. Zhang \textit{et al.} adopted 1D-Intersection over Union (IoU) as partially-spoofed audio detection evaluation\cite{9956134}. The intersection indicates the number of segments that are correctly predicted. The union is the sum of intersection and twice the number of segments that are mispredicted, as shown in Eq. \ref{eq:iou}.
    \begin{equation}
    \label{eq:iou}
    IoU=\frac{TP+TN}{TP+TN+2\times (FP+FN)}
    \end{equation}
     The system will be considered as a good detector if $IoU > \frac{1}{3}$.

     \item \textbf{Average precision (AP) and average recall (AR)}. AP measures the performance by averaging precision at different recall levels, providing a comprehensive assessment of precision and recall. 
     \begin{equation}
    \label{eq:ap}
    AP=\sum_{t}(R_t-R_{t-1})P_t
    \end{equation}
    where $R_t$ and $P_t$ are the recall and precision at the threshold \textit{t}. Usually, the threshold values are set at {0.5, 0.75, 0.9, 0.95}.
    
    AR focuses on the recall ability at different confidence thresholds, particularly useful in scenarios where high recall is essential, and average number of proposals \textit{N} are usually set to {5, 10, 20, 50, 100}. 
    \begin{equation}
    \label{eq:ar}
    AR=\frac{1}{N}\sum_{i=1}^NR(i)
    \end{equation}
\end{itemize}

\begin{table*}[h]
\caption{The existing competitions for partially deepfake audio task.(A: Audio, V: Video)}
\label{table:competition}
\centering
\begin{tabular}{cccccc}
\toprule[2pt]
 \rowcolor[HTML]{EFEFEF} 
Competitions                          & Track                          & Year & Mod. & Language     & 
\multicolumn{1}{c}{\cellcolor[HTML]{EFEFEF}URL} \\ \hline
ADD2022                               & Partially fake audio detection & 2022 & A    & Chinese      & 
\url{http://addchallenge.cn/add2022}
            \\
ADD2023                               & Manipulation region location   & 2023 & A    & Chinese      & 
\url{http://addchallenge.cn/add2023}
            \\
2024 1M-Deepfakes Detection Challenge & Deepfake Temporal Localization & 2024 & AV   & Multilingual & 
\url{https://deepfakes1m.github.io/2024}
        \\
2025 1M-Deepfakes Detection Challenge & Deepfake Temporal Localization & 2025 & AV   & Multilingual & 
\url{https://deepfakes1m.github.io/2025}        \\ \bottomrule[2pt]
\end{tabular}
\end{table*}

\subsection{Competitions}
In order to prosper this developing topic, some competitions have been organized to facilitate technical communication, summarizing in Table \ref{table:competition}.

The first Audio Deep Synthesis Detection Challenge (ADD2022)\footnote{ADD 2022: \url{http://addchallenge.cn/add2022}} is held in 2022, organized by Jianhua Tao and Haizhou Li\cite{conf/icassp/YiFTNMWWTBFLWZY22}. In this challenge, partially fake audio detection(PF) is firstly launched as an independent track, focusing on binary real/fake classification. Different from fully anti-spoofing task, it emphases on detecting the partially fake utterance with real or synthesized audio inserted from bona fide audio. EER is employed as the evaluation metric. The ADD 2022 is also launched as a Signal Processing Grand Challenge at the IEEE International Conference on Acoustics, Speech and Signal Processing in 2022 (ICASSP 2022). Additionally, based on this challenge, Tao \textit{et al.} also initiated a workshop on Deepfake Detection for Audio Multimedia at ACM Multimedia 2022 (DDAM 2022)\footnote{DDAM 2022: \url{http://addchallenge.cn/ddam2022}}. 

In 2023, the second Audio Deep Synthesis Detection Challenge (ADD 2023) is launched\footnote{ADD 2023: \url{http://addchallenge.cn/add2023}}. Different from ADD 2022, the setting of ADD 2023 goes beyond the goal of binary real/fake classification for entire utterances, which is the first competition focusing on localizing the manipulated intervals in partially fake audio (Track 2). The weighted sum of sentence-level Accuracy and segment-level $F_1$ is employed as the evaluation metric. The ADD 2023 challenge is also organized as part of the IJCAI 2023 Competitions and Challenges track, and the IJCAI 2023 Workshop on Deepfake Audio Detection and Analysis (DADA 2023) is organized based on it, leading widespread discussion within the scope. The systems on the leaderboard has become an important baseline for the following research in partially deepfake audio manipulation regions localization tasks.

The first 1M-Deepfakes detection challenge was launched in 2024, held by Abhinav Dhall \textit{et al.} at ACM Multimedia 2024\cite{cai20241m}. In this challenge, the \textit{Task2: Deepfake Temporal Localization} is to find out the timestamps [start, end] in which the manipulation is done, aiming at multi-modal data. AV-Deepfake1M Dataset is released in this challenge, which is multi-modality and multilingual. Recently, the second 1M-Deepfakes detection challenge is in progress. The new challenge is based on AV-Deepfake1M++ dataset containing over 2 million samples\cite{example}.

\section{Branches of Methods}
\begin{figure*}
    \centering
    \includegraphics[width=0.7\linewidth]{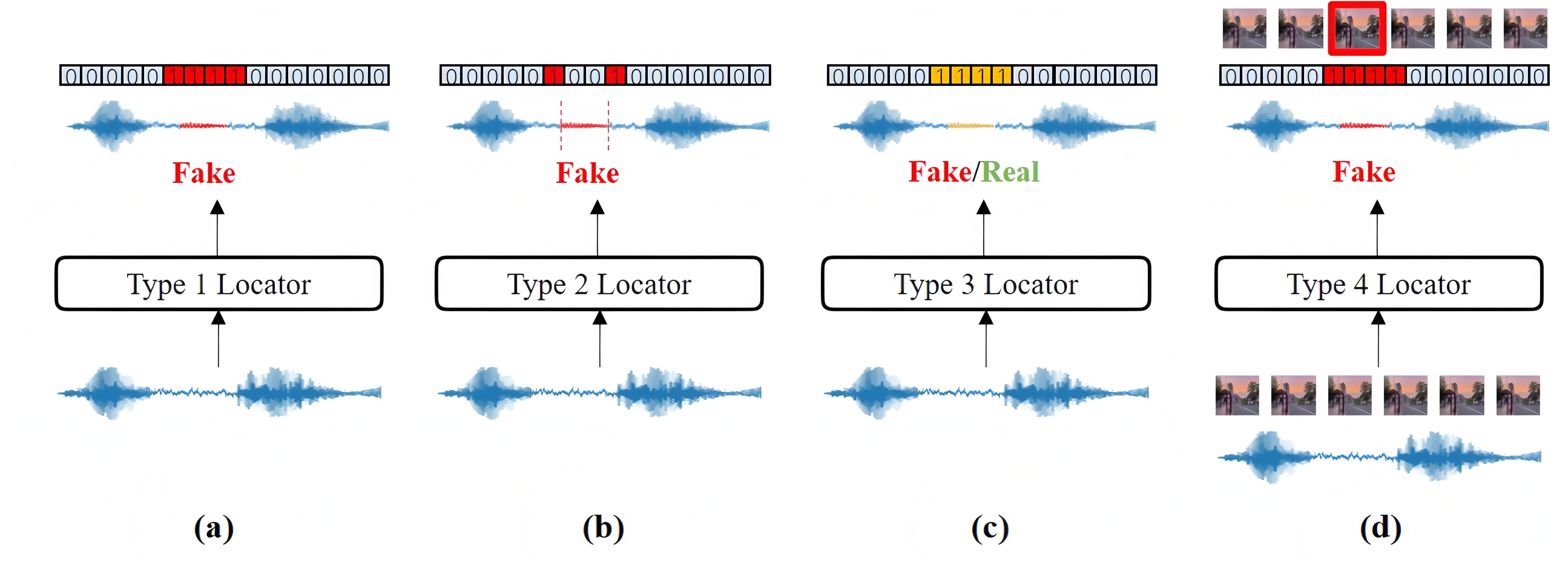}
    \caption{Four categories of locators in existing studies. (a) Locator based on frame-level authenticity; (b) Locator based on boundary perception; (c) Locator based on frame-level inconsistency; (d) Locator based on multi-modality fusion.}
    \label{fig:cate}
\end{figure*}
To date, there have been several studies working on partially deepfake audio manipulation region localization. All of these methods could be divided into FOUR types as shown in Fig.\ref{fig:cate}, and their strengths and weakness are summarized in Table \ref{tab:TYPES}.

\begin{table*}[]
\caption{The strengths and weakness for each type of method.}
\label{tab:TYPES}
\begin{tabular}{p{0.35cm}p{1.95cm}p{5.65cm}p{5.65cm}p{2.1cm}}
\toprule[2pt]
\rowcolor[HTML]{EFEFEF} 
Type & \makecell{Properties}               & \multicolumn{1}{c}{\cellcolor[HTML]{EFEFEF}Strengths}                                                                                                                                                  & \multicolumn{1}{c}{\cellcolor[HTML]{EFEFEF}Weakness}                                                                                & \makecell{Related methods} \\ \hline
\makecell{1}    & Frame-level Authenticity  & It is straightforward and constitutes the majority in existing research.                                                                                                                               & It may fail to locate the manipulation regions when the splicing clips are bona fide                                                & SPF\cite{cai2023dku}, TDL\cite{xie2024efficient}                \\
\rowcolor[HTML]{EFEFEF} 
\makecell{2}   & Boundary Perception       & It focuses on detecting stitching traces to avoid relying entirely on frame-level authenticity.                                                                                                        & It may fail when the splicing boundaries are hidden intentionally                                                                   & CFPRF\cite{10.1145/3664647.3680585}, BAM\cite{zhong2024enhancing}             \\
\makecell{3}    & Frame-level Inconsistency & It focuses on the inconsistency between frames instead of the authenticity, overcoming the weakness of the former two types. & For long lasting audio, the information in the utterances may change, the effectiveness needs further validation                    & PET\cite{10889913}, AGO\cite{10890470}, GNCL\cite{10888281}               \\
\rowcolor[HTML]{EFEFEF} 
\makecell{4}    & Multi-Modality Fusion     & It integrates multimodal forgery information and represents a new trend in recent research.                                                                                                            & It focuses more on the visual modality, and further explorations are needed in audio modality. & UMMAFormer\cite{10.1145/3581783.3613767}, W-TDL\cite{10.1145/3689092.3689410}      \\ \bottomrule[2pt]
\end{tabular}
\end{table*}

\subsection{Methods Based on Frame-level Authenticity }\label{sec:FLA}
For this category, the manipulation regions are detected based on the authenticity of segments. Due to the fact that commonly used datasets employ fake segments as the splicing clips, such as PartialSpoof, most existing methods belong to this type. Usually, two-stage frameworks are designed, consisting of a front-end feature extractor and a back-end classifier. MFCC\cite{5684887}, LFCC\cite{2018Integrated}, CQCC\cite{Todisco2016ANF}, Wav2vec\cite{2019arXiv190405862S,Tak2022AutomaticSV} and WavLM\cite{Chen_2022} are commonly employed as feature extractors while light convolutional neural network (LCNN)\cite{Wu2020LightCN}, ResNet\cite{2019arXiv190700501A}, SENet\cite{8701503} and long short-term memory (LSTM)\cite{6795963} are representative classifiers.

For example, Zhang \textit{et al.}\cite{9956134} propose to use CQCC and ANN as feature extractors and classifier respectively. A post-processing is employed to filter extreme short fake or real segments to modify the results. Zhang \textit{et al.}\cite{zhang2021multi} introduce a binary-branch multi-task models by integrating squeeze-and-excitation (SE) blocks with LCNN (SELCNN) and a BiLSTM to implement the basic model, employing LFCC as a front-end feature extractor. Zhu \textit{et al.}\cite{zhu2023local} add self-attention mechanism between SELCNN and BiLSTM to enhance the segment features, greatly improving the partially deepfake detection performance. Li \textit{et al.}\cite{li2023convolutional} adopt convolutional recurrent neural network (CRNN) to capture high temporal features and the context information. Li \textit{et al.}\cite{li2023multi} combine AASIST and Wav2Vec2 subsystems through multi-grained backend fusion to find out fake utterances or frames, where AASIST extracts features from utterance-level while Wav2Vec2 from segment-level. Martín-Doñas \textit{et al.}\cite{martin2023vicomtech} integrate Wav2Vec2 based feature extractor and BiLSTM to cluster the manipulated frames for partially deepfake detection.

Besides, some methods propose distinctive functional modules by combining these fundamental modules to enhance the performance. Xie \textit{et al.}\cite{xie2024efficient} propose temporal deepfake location (TDL) method to locate the manipulated regions. They devise an embedding similarity module to segregate authentic and synthetic frames within the embedding space to enhance the identification of genuine and fake distinctions at the embedding level. The result shows that it could achieve the EER at 7.04\% on PartialSpoof dataset at 160ms resolution, which was once the best performance of this dataset. Besides, it is also demonstrated that it could achieve the EER of 11.23\% on LAV-DF dataset, which reveals its good generalization ability. Inspired by this method, Dragar \textit{et al.}\cite{10.1145/3689092.3689410} modified TDL to a window-based method, named as W-TDL, and combined it with the EVA visual transformer to identify and localize manipulated segments in audio and visual data, achieving the best performance on AV-Deepfake1M dataset. Cai \textit{et al.}\cite{10.1016/j.csl.2023.101597,cai2023waveform,cai2023dku} designed the anti-spoofing detection system (SPF) to detect the fake frames embedding in the real audio. The Wav2Vec and WavLM are employed as feature extractors, and ResBlock is further used to learning the feature in-depth. Finally, transformer encoder and bidirectional long short term memory network (BiLSTM) are adopted as backend classifiers. It demonstrated that SPF achieves the champion of ADD 2023 Track 2 with the score of 0.6713.

Additionally, MUSAN noise\cite{2015arXiv151008484S}, reverberations and some other data augmentation strategies are usually employed to help enhance the robust of performance\cite{li2023convolutional,10.1016/j.csl.2023.101597,liu2023transsionadd,inbook}. Multi-domain feature fusion strategy has also been proposed\cite{10446471}.

However, although this type of methods dominates currently, it may fail to locate the manipulation regions when the splicing clips are bona fide.

\subsection{Methods Based on Boundary Perception }
For this category, the manipulation regions are detected via splicing traces. The intention is to focus on stitching traces and avoid relying entirely on fragment-level authenticity. The study shows that the partially spoofed audio-trained CMs significantly focus on the transition regions created by the overlap-add operation during the dataset creation\cite{liu2024neural}. However, the biggest obstacle encountered by such methods is data bias. Thus, some of these existing methods appear simultaneously with frame-level authenticity classifiers.

Wu \textit{et al.}\cite{9746162} introduce a question-answering (QA) strategy based on SE-ResNet architecture with self-attention mechanism to locate the manipulated regions by predicting the start and end positions of clips. Zeng \textit{et al.}\cite{10037855} adopt a ResNet-based model for splicing traces localization, both time and frequency features are considered. The localization probability are obtained via 4 consecutive frames. They conducted experiments on their own dataset and achieved the $F_1$ at 0.741 in the test set with chunk size of 64 frames. Cai \textit{et al.}\cite{10.1016/j.csl.2023.101597,cai2023waveform,cai2023dku} proposed a dual-head system to detect the a frame-level anti-spoofing and locate the boundary simultaneously. Boundary detection system (BDR) is designed to perceive the splicing boundaries with the same network structure as SPF. It is claimed to achieve the frame-level EER at 0.064\% on ADD2023 Track 2 dev set with training on ADD2023 Track 2 train set and at 1.74\% on PartialSpoof eval set with training on PartialSpoof train set for boundary frames detection. Wu \textit{et al.}\cite{10.1145/3664647.3680585} introduced a coarse-to-fine proposal refinement framework (CFPRF) to locate the partially fake. They proposed the temporal forgery localization (TFL) network to predict the precise timestamps at which these forgery segments start and end. It reveals that CFPRF could achieve the EER at 0.08 on HAD dataset, 7.41 on PartialSpoof dataset, and 0.82 on LAV-DF dataset for fake segments localization, which are claimed to be superior to the method mentioned in Ref.\cite{10.1016/j.csl.2023.101597} and \cite{Zhang_2023}. Zhong \textit{et al.}\cite{zhong2024enhancing} proposed boundary-aware attention Mechanism (BAM), consisting of boundary enhancement (BE) module and boundary frame-wise attention (BFA) module, to improve the accuracy and localization capability by using boundary information. BE aims to extract intra-frame and inter-frame information to enhance boundary features for splicing boundary detection and authenticity detection. BFA aims to use boundary prediction results to explicitly control the feature interaction between frames, in order to effectively distinguish between real and fake frames. When using WavLM as the front-end feature, the BAM method ontained an EER of 3.58\% at a resolution of 160ms, achieving the state-of-the-art performance on the PartialSpoof dataset.

Obviously, the existing methods inherit the drawbacks of the first type of method if they partially relied on the frame-level authenticity. Besides, facing with the increasingly advanced splicing technology, in real adversarial attack scenarios, the splicing boundaries will be intentionally hidden, and such methods are prone to failure.

\subsection{Methods Based on Frame-level Inconsistency }
For this category, the manipulation regions are detected based on the inconsistency between the manipulated regions and non-manipulated ones, which can provide effective solutions to overcome the difficulties encountered by the two types mentioned above. Existing studies indicate that the systems composed of frame-wise consistency related modules usually exhibit superior performance. According to the existing methods, there are three subtypes:
\subsubsection{Difference-Aware Between Real and Fake Frames}
In CFPRF\cite{10.1145/3664647.3680585}, Difference-Aware Feature Learning Module (DAFL) is proposed to enhance the difference between real and fake frames. Also, In TDL\cite{xie2024efficient}, the embedding similarity module is designed to capture the differences in feature learning between real and fake frames and employed as mask to enhance the diversity. The designation of BE module in BAM\cite{zhong2024enhancing} is also the same.

\subsubsection{Distribution Shift Between Manipulated and Non-manipulated Regions} Zeng \textit{et al.}\cite{10890470} proposed the adversarial training and gradient optimization (AGO) method to locate the partially deepfake segments by focusing on the distribution shift between manipulated and non-manipulated regions, which provides a new perspective to address the issue. Gradient reversal layer (GRL) is employed to reduce the dependence of model on specific domain features and enhance the generalization ability. The results show that AGO could achieve the segment-level $F_1$ score at 0.7187 and ADD2023 score at 0.8254 on ADD2023 Track 2 dataset, which is a relative improvement of 22.82\% than SPF\cite{10.1016/j.csl.2023.101597,cai2023waveform,cai2023dku} without any data augmentation strategies. In PartialSpoof dataset, it could achieve the EER at 6.79, which is superior than that of CFPRF\cite{10.1145/3664647.3680585} at 7.41.
\begin{table*}[h]
\caption{Comparison of prominent methods in partially deepfake audio localization tasks. The metrics mentioned in the table are all in segment-level. The results are all from citations. (The "*" indicates that the CFPRF method was trained on the training set of the LAV-DF dataset and tested on its test set, while the results of other methods on LAV-DF are trained with PartialSpoof train set.)}
\centering
\label{tab:com1}
\begin{tabular}{cccccccc}
\toprule[2pt]
\rowcolor[HTML]{EFEFEF} 
\multicolumn{1}{c}{\cellcolor[HTML]{EFEFEF}}                        & \cellcolor[HTML]{EFEFEF}                       & \multicolumn{2}{c}{\cellcolor[HTML]{EFEFEF}PartialSpoof} & \multicolumn{2}{c}{\cellcolor[HTML]{EFEFEF}ADD2023 Track 2} & \multicolumn{2}{c}{\cellcolor[HTML]{EFEFEF}LAV-DF} \\ \cline{3-8} 
\rowcolor[HTML]{EFEFEF} 
\multicolumn{1}{c}{\multirow{-2}{*}{\cellcolor[HTML]{EFEFEF}Model}} & \multirow{-2}{*}{\cellcolor[HTML]{EFEFEF}Year}  & EER($\%$)$\downarrow$ /Resolution  & $F_1$$\uparrow$     & EER($\%$)$\downarrow$   & $F_1$$\uparrow$ /Resolution       & EER($\%$)$\downarrow$  & $F_1$$\uparrow$     \\ \hline
LCNN-BLSTM(w LFCC)\cite{2021An,xie2024efficient}                  & 2021                  & 16.21 / 160ms          & -               & -               & -                 & 17.89          & 0.8338          \\
LCNN-BLSTM(w W2V2-XLS-R)\cite{2021An,xie2024efficient}            & 2021                  & 9.87 / 160ms           & -               & -               & -                 & 15.35          & 0.7650          \\
SELCNN-BLSTM\cite{zhang2021multi}                                 & 2021                  & 15.93 / 160ms          & -               & -               & -                 & -              & -               \\
SPF(w WavLM)\cite{10.1016/j.csl.2023.101597,cai2023dku}         & 2023                  & -              & 0.9296          & -               & 0.6066 / 20ms            & -              & -               \\
TranssionADD\cite{liu2023transsionadd,yi2024add2023audiodeepfake} & 2023                  & -              & -               & -               & 0.5460 / 160ms            & -              & -               \\
CRNN\cite{li2023convolutional}                                    & 2023                  & -              & -               & -               & 0.5449 / 10ms            & -              & -               \\
Multi-grained Backend Fusion\cite{li2023multi}                    & 2023                  & -              & -               & -               & 0.5253 / 20ms            & -              & -               \\
Vicomtech\cite{martin2023vicomtech}                               & 2023                  & -              & -               & -               & 0.5167 / 20ms            & -              & -               \\
TDL\cite{xie2024efficient}                                        & 2024                  & 7.04 / 160ms           & -               & -               & -                 & \textbf{11.23} & \textbf{0.8551} \\
BAM(w WavLM-Large)\cite{zhong2024enhancing}                       & 2024                  & \textbf{3.58 / 160ms}  & \textbf{0.9609} & -               & -                 & -              & -               \\
CFPRF\cite{10.1145/3664647.3680585}                               & 2024                  & 7.41 / Not found           & 0.9389          & -               & -                 &\textbf{ 0.82* }         & \textbf{0.9956* }        \\
AGO\cite{10890470}                                                & 2025                  & 6.79 / 40ms           & 0.9436          & -               & 0.7187 / 40ms            & -              & -               \\
GNCL\cite{10888281}                                               & 2025                  & 11.81 / 20ms          & 0.8979          & -               & -                 & -              & -               \\
PET\cite{10889913}                                                & 2025                  & -              & -               & 29.50           & \textbf{0.7397 / 10ms}   & -              & -               \\ \bottomrule[2pt]
\end{tabular}
\end{table*}
\begin{table*}[h]
\caption{Comparison of prominent methods in partially deepfake audio-video localization tasks. The results are all from citations.}
\label{tab:com2}
\begin{tabular}{>{\centering\arraybackslash}p{2.7cm}cccccccccccc}
\toprule[2pt]
\rowcolor[HTML]{EFEFEF} 
Model                                                            & Dataset                         & Year & Mod. & AP@0.95        & AP@0.9                                               & AP@0.75        & AP@0.5         & AR@5           & AR@10          & AR@20          & AR@50          \\ \hline
AVFusion\cite{bagchi2021hear}                                    &                                 & 2021 & AV   & 0.11           & -                                                    & 23.89          & 65.38          & -              & 62.98          & 59.26          & 54.80          \\
BA-TFD\cite{10034605}                                            &                                 & 2022 & AV   & 0.29          & \textbf{-}                                           & 47.06          & 76.90          & \textbf{-}     & 59.32          & 61.19          & 64.52          \\
BA-TFD+\cite{CAI2023103818,10.1145/3689092.3689410}              &                                 & 2023 & AV   & 4.44          & -                                                    & 84.96          & 96.30          & -              & 78.75          & 79.40          & 80.48          \\
UMMAFormer\cite{10.1145/3581783.3613767,10.1145/3689092.3689410} &                                 & 2023 & AV   & 37.61          & -                                                    & \textbf{95.54} & \textbf{98.83} & -              & 92.10          & 92.42          & \textbf{92.48} \\
CFPRF\cite{10.1145/3664647.3680585}                              & \multirow{-5}{*}{LAV-DF}        & 2024 & A    & \textbf{88.64} & \textbf{91.65}                                       & 93.47          & 94.52          & \textbf{93.51} & \textbf{93.51} & \textbf{93.51} & -              \\ \hline
BA-TFD\cite{10034605}                                            &                                 & 2022 & AV   & 0.02          &0.19 & 6.34          & 37.37          & 26.82          & 30.66          & 35.95          & 45.55          \\
BA-TFD+\cite{CAI2023103818,10.1145/3689092.3689410}              &                                 & 2023 & AV   & 0.03          & 0.48                                                & 13.64          & 44.42          & 29.88          & 34.67          & 40.37          & 48.86          \\
UMMAFormer\cite{10.1145/3581783.3613767,10.1145/3689092.3689410} &                                 & 2023 & AV   & 1.58          & 07.65                                                & 28.07          & 51.64          & 40.27          & 42.09          & 43.45          & 44.07          \\
W-TDL\cite{10.1145/3689092.3689410}                        & \multirow{-4}{*}{AV-Deepfake1M} & 2024 & AV   & \textbf{50.66} & \textbf{70.43}                                       & \textbf{88.75} & \textbf{94.75} & \textbf{88.78} & \textbf{89.13} & \textbf{89.17} & \textbf{89.17} \\ \bottomrule[2pt]
\end{tabular}
\end{table*}

\subsubsection{Inconsistency Between Manipulated and Non-manipulated Regions} He \textit{et al.}\cite{10889913} initialed a partially deepfake audio localization method via empirical wavelet transform and temporal self-consistency learning (PET), locating manipulated regions via temporal self-consistency learning of high-frequency components. Different from existing methods, PET directly utilizes the frame-wise similarity of high-frequency components as a feature to capture the inconsistency among frames. It is a location-only system that could achieve the state-of-the-art segment-level $F_1$ score at 0.7397, 2.92\% relative improvement compared to AGO\cite{10890470} and 21.94\% higher than that of SPF\cite{10.1016/j.csl.2023.101597,cai2023waveform,cai2023dku} at 0.6066. Ge \textit{et al.}\cite{10888281} proposed a graph neural network with consistency
loss (GNCL) to locate the spoofed segments. The consistency-enhanced loss function is introduced to bridge different. It achieves the EER at 11.81\% on PartialSpoof dataset at a 20ms resolution.

However, although these methods have achieved good results for audio clips with a few seconds long,  their effectiveness for longer lasting audio, such as continuous recordings spanning several hours or more, needs further validation.

\subsection{Methods Based on Multi-Modality Fusion} AVFusion\cite{bagchi2021hear} is the first model to jointly
consider audio and video modalities for temporal action localization, aiming to locate the start and end timestamps of activities in the video stream. Based on that, some studies\cite{10.1145/3394171.3413700,10034605} are initialed for temporal forgery localization to locate the start and end timestamps of manipulated segments. Cai \textit{et al.}\cite{10034605} proposed BA-TFD and BA-TFD+, two multi-modality methods, for content-driven partially deepfake audio-video detection and illustrated its effectiveness on LAV-DF dataset. They are now also considered as baseline methods on the LAV-DF dataset. Zhang \textit{et al.}\cite{10.1145/3581783.3613767} proposed UMMAFormer to predict forgery segments and their corresponding start and end timestamps in untrimmed videos or audios, considering three scenarios: visual-only, audio-only, and joint audio-visual modalities. In UMMAFormer, a Temporal Feature Abnormal
Attention (TFAA) module is built from reconstruction learning and Cross-Reconstruction Attention Transformer (CRATrans) block to identify abnormal segments. The results reported that, compared to BA-TFD, the AP@0.5
has increased from 76.90\% to 98.83\%, and from 0.29\% to 37.61\% at AP@0.95 on LAV-DF dataset. Further more,  CFPRF\cite{10.1145/3664647.3680585} refreshed the AP@0.95 to 88.61\%. Besides,  Cai \textit{et al.} further expanded the LAV-DF dataset to the AV-Deepfake1M dataset\cite{10.1145/3664647.3680795}, and organized the 2024 1M-Deepfakes Detection Challenge on ACM Multimedia 2024\footnote{2024 1M-Deepfakes Detection Challenge: \url{https://deepfakes1m.github.io/2024}}. In the challenge, W-TDL, proposed by Dragar \textit{et al.}\cite{10.1145/3689092.3689410}, is confirmed to outperform existing state-of-the-art techniques on the AV-Deepfake1M dataset.

However, currently this type of method focuses more on the aspect of visual modality, and further explorations are needed in audio modality.

Additionally, beyond these types, the audio copy-move forgery detection task also can be considered as a simplified situation of this manipulation regions localization tasks, which is a legacy technique with the copied frames being selected from the audio itself and then being inserted or replaced at certain position in the audio\cite{imran2017blind}. The pipeline of solving audio copy-move forgery detection usually begins with voice activity detection module(VAD). Then discrete cosine transform(DCT) coefficients, the constant Q spectral sketches (CQSS), discrete Fourier transform (DFT), MFCC, etc. are employed as feature extractor to obtain the feature representation for segments. Euclidean distance(ED), dynamic time warping(DTW) and cosine similarity are adopted as measurement to calculate pairwise distance or similarity between segments in order to locate the copy-move forgery regions\cite{9921343, li2019homologous, akdeniz2022linear, ustubioglu2022robust, akdeniz2024detecting, ustubioglu2023detection, xie2018copy, su2023robust, wang2017algorithm, zhao2024audio}. Obviously, methods for audio copy-move forgery detection are mainly based on the pairwise similarity of waveform or spectrum between clips, which will have limitations when used for partially deepfake manipulation regions localization generated via cutting-edge techniques. But they can inspire us to achieve the goal by constructing deeper feature for localization via frame-wise similarities.

\section{Comparisons of Existing Methods}\label{sec:com}
In this section, the comparisons of some methods that utilize common datasets are demonstrated, including PartialSpoof, ADD2023 Track 2 dataset, LAV-DF and AV-Deepfake1M dataset. There are mainly two groups for comparison, one for audio-only (See Table \ref{tab:com1}) and another for audio-video datasets (See Table \ref{tab:com2}).

In Table \ref{tab:com1}, it reveals the significant technological improvements in partially deepfake audio localization. The results show that, in the past five years since the issue was raised, the segment-level EER of the PartialSpoof dataset has decreased from 16.21\% to 3.58\%(BAM), and the segment-level $F_1$ has increased to 0.9609. For ADD2023 Track 2 dataset, the segment-level $F_1$ has increased from 0.6066 to 0.7397(PET). CFPRF, as a uni-modality model, could achieve the segment-level EER at 0.82\%, demonstrating the potential ability of partially deepfake audio localization methods in multi-modality partially deepfake datasets. Some methods also show well cross-domain localization capabilities. Specifically, TDL could achieve the segment-level EER at 11.23\% on LAV-DF dataset while training on the PartialSpoof train set.

Table \ref{tab:com2} shows the comparison of partially deepfake localization models for audio-video datasets, illustrating remarkable progress in multi-modality partially deepfake localization tasks that incorporating audio as one of the key modalities. The latest research shows that, for the LAV-DF dataset, the AP@0.95 has soared from 0.29(BA-TFD) to 88.64(CFPRF) and the AP@0.5 has increased from 76.90(BA-TFD) to 98.83(UMMAFormer). The scores of each item for average recall(AR) have approximate 50\% relative improvements. For AV-Deepfake1M, the AP@0.95 has soared from 0.02(BA-TFD) to 50.66(W-TDL) and the AP@0.5 has increased from 37.37(BA-TFD) to 94.97(W-TDL). The score of AR@5 surges from 26.82(BA-TFD) to 88.78(W-TDL) and AR@50 has achieved a relative improvement of approximately 96\%. Besides, the superior performance of CFPRF and W-TDL emphasizes the potential benefits of partially deepfake audio localization methods in multi-modality partially deepfake localization tasks that incorporating audio as one of the key modalities.

\section{Challenges and Development Trends}
\subsection{Challenges and Limitations} 
\subsubsection{\textbf{Insufficient localization accuracy}} According to Sec.\ref{sec:com}, although some significant improvements have achieved, however, as shown in Table \ref{tab:com2}, there is still a long way to go for lower resolution and complex situations, such as audio with "truth for truth" manipulation, smoothing processing for splicing boundaries, etc.. Besides, the distribution shift for manipulated clips and environment shift for long lasting audio are also needed to be taken into considerations.

\subsubsection{\textbf{Lack of evidence to support the results}} The existing methods typically use binary sequences to indicate the location of the manipulations, answering the question of 'what' but lacking a response to 'why'. Specifically, in practical applications, more detailed physical evidence is needed to support the results, such as some indicators or measurements refer to spectral discontinuities, changes in timbre, phase information missing, etc.

\subsection{Potential Trends}
\subsubsection{\textbf{Focusing on the inconsistency between manipulated and non-manipulated regions instead of their authenticity}} Manipulations with real clips is very likely to occur in real-world scenarios, so in order to improve the practical ability and enhance the generalization capability, researchers need to shift their attentions to the essential features that are highly relevant to the differences between manipulated and non-manipulated regions, such as acoustic feature distribution shifting, noise inconsistency, sound field inconsistency, emotional inconsistency, etc..
\subsubsection{\textbf{Utilizing LLM-based methods to provide the evidence}} Regarding the issue of lacking physical evidence, LLM may be helpful. There have been some studies on the application of speech LLM, but there has been no breakthrough in partially deepfake audio localization and its forensics. Researchers may further construct indicators related to the physical evidence and take full advantage of LLM's reasoning capabilities to obtain the physical evidence.
\subsubsection{\textbf{Expanding to multi-modality deepfake localization tasks}} Given the newly proposed challenge competitions and datasets, the trend of multi-modal partially deepfake localization has emerged, but according to the studies, the current multi-modal deepfake localization evaluation usually focuses on visual modality, maybe in the near future, the audio part will play an important role.

\section{Conclusions}
In this survey, we sort out the route to development of partially deepfake localization tasks, including datasets, evaluation metrics, challenge competitions, branches of existing methods, current limitations and potential trends, providing a comprehensive insight of this scope for beginners to catch up with. Specifically, we elaborate on the definition of partially deepfake audio localization and sorted out the current research status, including the method with the best performance on diverse datasets. Based on the achievements already made, potential trends of this scope is discussed. We hope this survey could be the reference for later researchers and bring about deeper thought and exploration.

\section*{Acknowledgments}
This work is supported by the National Natural Science Foundation of China (NSFC) (No.62206278, No. 62322120, No. 62306316).

\bibliographystyle{IEEEtran}
\bibliography{bibfile}
%


 

\begin{IEEEbiography}[{\includegraphics[width=1in,height=1.25in,clip,keepaspectratio]{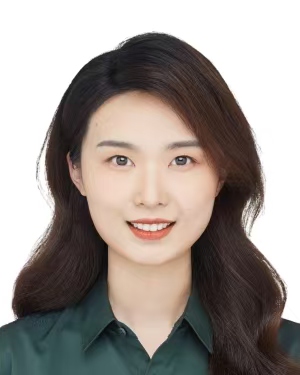}}]{Jiayi He} received the Ph.D. degree from Beijing Jiaotong University, Beijing, China, in 2021, and the B.S. degree from Wuhan University of Technology, Wuhan, China, in 2016. From 2019 to 2020, she was a visiting student at Harvard University, Boston, MA, USA. She is currently an Assistant Researcher with the State Key Laboratory of Multimodal Artificial Intelligence Systems, Institute of Automation, Chinese Academy of Sciences, Beijing, China. Her research interests include fake audio detection and audio forensics.
\end{IEEEbiography}
\begin{IEEEbiography}[{\includegraphics[width=1in,height=1.25in,clip,keepaspectratio]{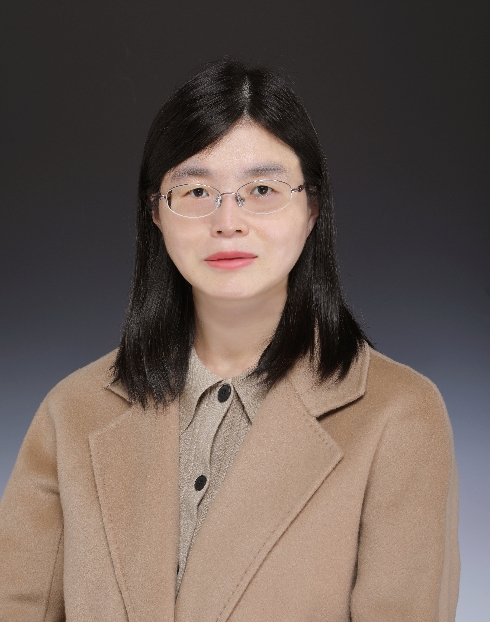}}]{Jiangyan Yi} (Member, IEEE) received the Ph.D. degree from the University of Chinese Academy of Sciences, Beijing, China, in 2018, and the M.A. degree from the Graduate School of Chinese Academy of Social Sciences, Beijing, China, in 2010. From 2011 to 2014, she was a Senior R\&D Engineer with Alibaba Group. From 2018 to 2024, she was an Associate Professor with the State Key Laboratory of Multimodal Artificial Intelligence Systems, Institute of Automation, Chinese Academy of Sciences, Beijing, China. She is currently an Associate Researcher with the Department of Automation, Tsinghua University. Her research interests include speech signal processing, speech recognition and synthesis, fake audio detection, audio forensics, and transfer learning.
\end{IEEEbiography}
\begin{IEEEbiography}[{\includegraphics[width=1in,height=1.25in,clip,keepaspectratio]{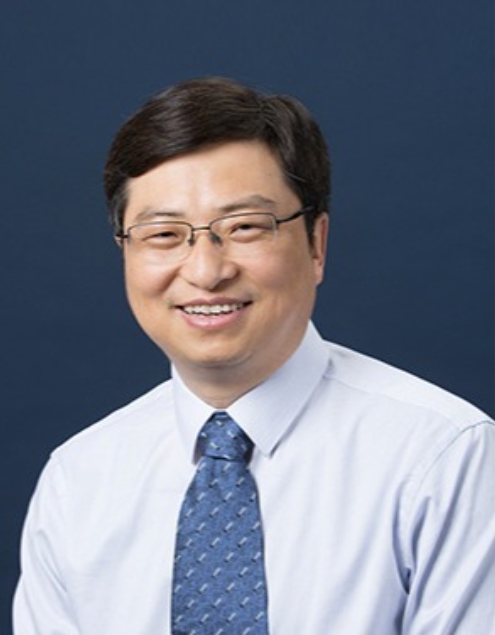}}]{Jianhua Tao} (Senior Member, IEEE)  received the M.S. degree from Nanjing University, Nanjing, China, in 1996, and the Ph.D. degree from Tsinghua University, Beijing, China, in 2001. He is currently a Professor with Department of Automation, Tsinghua University, Beijing, China. He has authored or coauthored more than 300 papers on major journals and proceedings including the IEEE TASLP, IEEE TAFFC, IEEE TIP, IEEE TSMCB, Information Fusion, etc. His current research interests
include speech recognition and synthesis, affective computing, and pattern recognition. He is the Board Member of ISCA, the chairperson of ISCA SIG-CSLP, the Chair or Program Committee Member for several major conferences, including Interspeech, ICPR, ACII, ICMI, ISCSLP, etc. He was the subject editor for the Speech Communication, and is an Associate Editor for Journal on Multimodal User Interface and International Journal on Synthetic Emotions. He was the recipient of several awards from important conferences, including Interspeech, NCMMSC, etc. 
\end{IEEEbiography}

\begin{IEEEbiography}[{\includegraphics[width=1in,height=1.25in,clip,keepaspectratio]{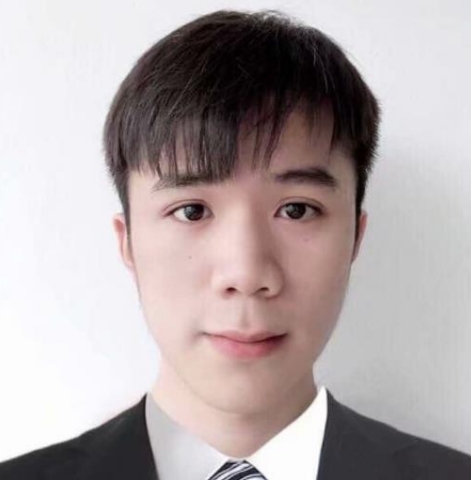}}]{Siding Zeng} received the B.E. degree from Sichuan Agricultural University, Chengdu, China, in 2021. He is currently a Master’s candidate jointly supervised by the University of Chinese Academy of Sciences and State Key Laboratory of Multimodal Artificial Intelligence Systems, Institute of Automation, Chinese Academy of Sciences, Beijing, China. His research interests include audio deepfake detection, unsupervised domain adaptation, and multimodal learning.
\end{IEEEbiography}

\begin{IEEEbiography}[{\includegraphics[width=1in,height=1.25in,clip,keepaspectratio]{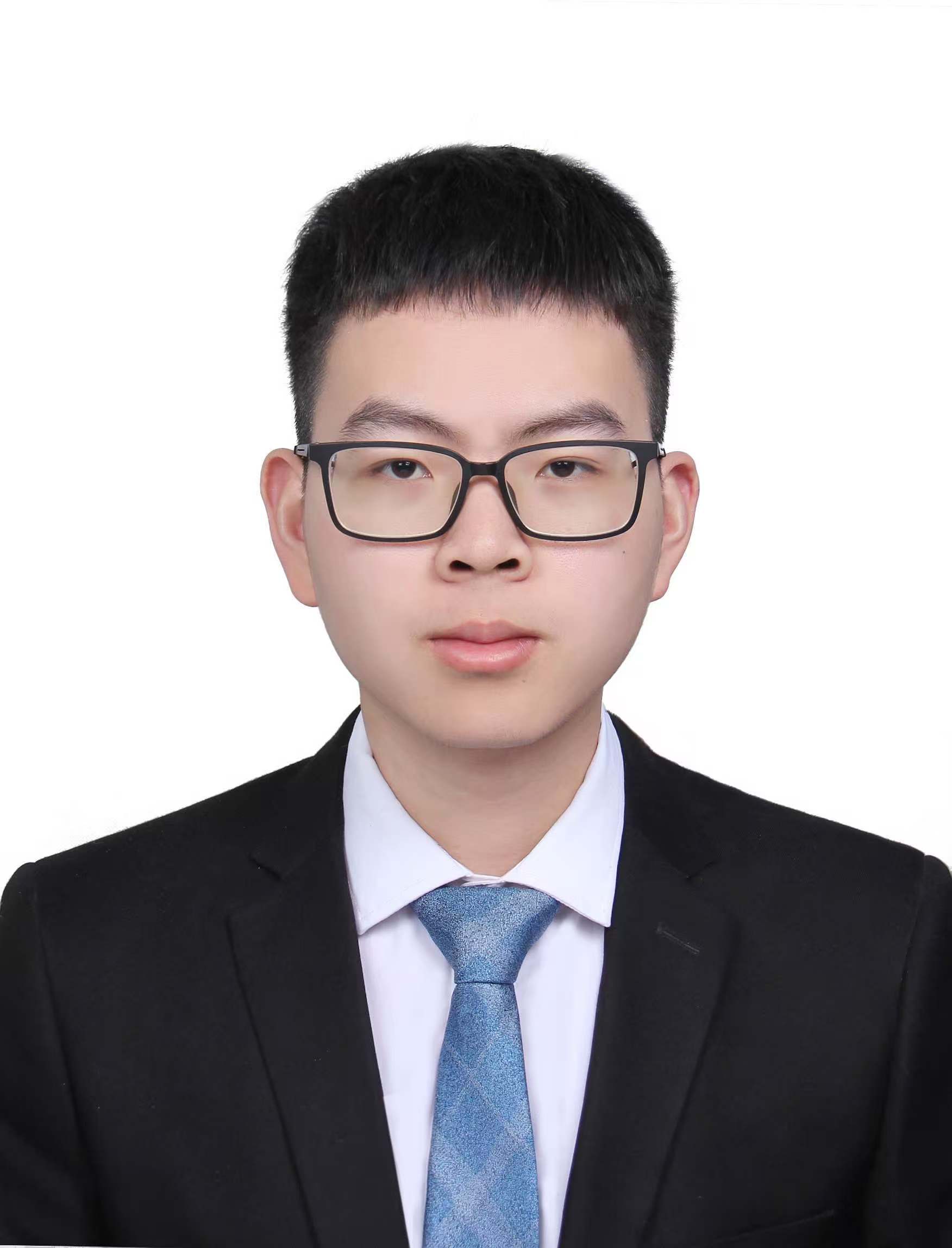}}]{Hao Gu} received the B.S. degree from Harbin Institute of Technology, Harbin, China, in 2022. He is currently a Ph.D. candidate at State Key Laboratory of Multimodal Artificial Intelligence Systems, Institute of Automation, Chinese Academy of Sciences, Beijing, China. His current research interest include fake audio detection.
\end{IEEEbiography}



\vfill

\end{document}